\documentclass[12pt,preprint]{aastex}

\shorttitle{The Fermi Bubbles}
\shortauthors{Fujita et al.}

\begin{document}


\title{The Fermi Bubbles as a Scaled-up Version of Supernova Remnants}

\author{Yutaka Fujita}
\affil{Department of Earth and Space Science, Graduate School of
 Science, Osaka University, Toyonaka, Osaka 560-0043, Japan}
\email{fujita@vega.ess.sci.osaka-u.ac.jp}

\and

\author{Yutaka Ohira and Ryo Yamazaki }
\affil{Department of
Physics and Mathematics, Aoyama Gakuin University, Fuchinobe, Chuou-ku,
Sagamihara 252-5258, Japan}

\begin{abstract}
 In this study, we treat the Fermi bubbles as a scaled-up version of
 supernova remnants (SNRs). The bubbles are created through activities
 of the super-massive black hole (SMBH) or starbursts at the Galactic
 center (GC). Cosmic-rays (CRs) are accelerated at the forward shocks of
 the bubbles like SNRs, which means that we cannot decide whether
 the bubbles were created by the SMBH or starbursts from the radiation
 from the CRs. We follow the evolution of CR distribution by solving a
 diffusion-advection equation, considering the reduction of the
 diffusion coefficient by CR streaming. In this model, gamma-rays are
 created through hadronic interaction between CR protons and the gas in
 the Galactic halo. In the GeV band, we can well reproduce the observed
 flat distribution of gamma-ray surface brightness, because some amount
 of gas is left behind the shock. The edge of the bubbles is fairly
 sharp owing to the high gas density behind the shock and the reduction
 of the diffusion coefficient there. The latter also contributes the
 hard gamma-ray spectrum of the bubbles. We find that the CR
 acceleration at the shock has started when the bubbles were small, and
 the time-scale of the energy injection at the GC was much smaller than
 the age of the bubbles. We predict that if CRs are accelerated to the
 TeV regime, the apparent bubble size should be larger in the TeV band,
 which could be used to discriminate our hadronic model from other
 leptonic models. We also present neutrino fluxes.
\end{abstract}


\section{Introduction}

The Fermi bubbles are huge gamma-ray bubbles discovered with {\it Fermi
Gamma-ray Space Telescope} in the direction of the Galactic center (GC)
in the GeV band (\citealt*{su10a}; see also \citealt{dob10}). They are
symmetric about the Galactic plane and the size is $\sim 50^\circ$
($\sim 10$~kpc). Their surface brightness is relatively uniform, and they
have sharp edges and hard spectrum \citep{su10a}.

Several models have been proposed for the origin of the bubbles. These
models assume that activities of the super-massive black hole (SMBH) or
starbursts at the GC created the bubbles. Some models indicated that
cosmic-rays (CR) that are accelerated via star formation activities are
conveyed into the bubbles \citep{cro11a,cro12a}. Others pointed out that
the CRs are originated from jets launched by the central black hole
\citep{guo12a}, or accelerated inside the bubbles
\citep{mer11a,che11a}. Gamma-rays can be generated by interaction
between CR protons and ambient gas (hadronic models), or by inverse
Compton scattering by CR electrons (leptonic models).

In this study, we treat the Fermi bubbles as a scaled-up version of
supernova remnants (SNRs). CRs are accelerated at the forward shock
front like SNRs. We explicitly solve a diffusion-advection equation to
study the evolution of CR distribution. We also focus on the reduction
of the diffusion coefficient around the bubbles, which slows CR
diffusion and is crucial to explain the sharp edge of the bubbles
\citep{guo12b}. The reduction has been indicated and studied for SNRs
\citep*{tor08b,fuj09c,li10a,ohi11b,fuj11a,yan12b,nav13a,mar13a}; it
could be caused by a CR streaming instability or anisotropic
diffusion. For the Fermi bubbles, \citet{yan12a} studied the reduction
by the latter. In this study, we investigate the former. We refer to
protons as CRs unless otherwise mentioned.

\section{Models}

For the sake of simplicity, we assume a spherically symmetric bubble,
and mainly focus on the high-galactic-latitude part of the Fermi bubbles
(large $|b|$ and small $|l|$ in the Galactic coordinate). Before the
bubble is born, the gas in the Galactic halo is static and has a
distribution of $\rho_0(r)=\rho_1(r/r_1)^{-\omega}$, where $r$ is the
distance from the GC, and $\rho_1$, $r_1$, and $\omega$ are the
parameters. We assume that the adiabatic index of the gas is
$\gamma=5/3$. For hydrodynamic evolution of the halo gas, we adopt the
Sedov-Taylor solution \citep[e.g.][]{mih84,ost88}. If an energy is
injected at $t=0$ at the GC, the radius of the shock front of the bubble
can be written as
\begin{equation}
\label{eq:Rs}
 R_{\rm sh}(t) = \xi\left(\frac{E_{\rm tot}}
{\rho_1 r_1^\omega}\right)^{1/(5-\omega)}
t^{2/(5-\omega)}\;,
\end{equation}
where $\xi\sim 1$, and $E_{\rm tot}$ is the injected energy.  On the
other hand, if the energy is continuously injected at a rate of $L_w$ at
the GC, it should be
\begin{equation}
\label{eq:Rsw}
 R_{\rm sh}(t) = \xi\left(\frac{L_w}
{\rho_1 r_1^\omega}\right)^{1/(5-\omega)}
t^{3/(5-\omega)}\;.
\end{equation}
We ignore the effect of CR pressure on the gas. We do not care about the
energy source: it may be the SMBH or starburst activities at the GC. If
the gas has a finite temperature, the Mach number of the shock gradually
decreases as the velocity, $V_{\rm sh}=dR_{\rm sh}/dt$, decreases. The
gas density $\rho$ and velocity $u$ for $r<R_{\rm sh}$ follow the
Sedov-Taylor solution.

Our CR model is based on the one in \citet*{fuj10a} for the evolution of
SNRs. However, while we adopted a Monte Carlo approach in \citet{fuj10a}
to calculate the CR distribution, in this study we explicitly solve a
diffusion-advection equation to follow the evolution of a CR
distribution function $f(r,p,t)$, where $p$ is the momentum of CRs. The
equation is
\begin{equation}
\label{eq:diff}
 \frac{\partial f}{\partial t} 
= \frac{1}{r^2}\frac{\partial}{\partial r}\left(r^2 \kappa\frac{\partial
					       f}{\partial r}\right)
- (u+u_w)\frac{\partial f}{\partial r}
+ \frac{1}{3 r^2}\left[\frac{\partial}{\partial r}(r^2 (u+u_w))\right]
p\frac{\partial
f}{\partial p} + Q\:,
\end{equation}
where the source $Q$ describes particle injection, and $u_w$ is the
velocity of the waves that scatter CRs. We assume that $u_w=v_A$ for
$r>R_{\rm sh}$, where $v_A$ is the Alfv\'en velocity, and that $u_w=0$
for $r<R_{\rm sh}$ because the waves would isotropically propagate for
$r<R_{\rm sh}$.

CRs are accelerated at the shock front of the bubble ($r=R_{\rm
sh}$). We do not consider the details of particle acceleration. CRs are
accelerated in the shock neighborhood, where some nonlinear effects
generate strong magnetic waves or cause strong amplification of magnetic
fields \citep{luc00a,bel04a}. In this region, particle diffusion would
follow the so-called Bohm diffusion and CR acceleration is
effective. The spatial scale of the region is much smaller than $R_{\rm
sh}$ and we ignore the width. Thus, we assume that
\begin{equation}
\label{eq:Q}
 Q(r,p,t)=\left\{\begin{array}{ll}
		 K_Q^{-1} p^{-q} \rho(R_{\rm sh,+})V_{\rm sh}^3
\delta(r-R_{\rm sh}) & \mbox{if $p_{\rm min}<p<p_{\rm max}$} \\
0 & \mbox{otherwise,}
		      \end{array}
\right.
\end{equation}
where $q$ is the parameter, and $R_{\rm sh,+}$ is the radius just
outside the shock. At a strong shock, the standard diffusive shock
acceleration model predicts that $q\sim 4$ \citep{dru83a}. For instant
energy injection (equation~(\ref{eq:Rs})), the coefficient is written as
\begin{equation}
 K_Q = 16\pi^2 c\: \xi^{5-\omega}\left(\frac{2}{5-\omega}\right)^3
\frac{E_{\rm tot}}{E_{\rm cr,tot}}\ln\left(\frac{t_f}{t_0}\right)
\int_{p_{\rm min}}^{p_{\rm max}}p'^{\: 2-q}
\sqrt{p'^{\: 2}+m_p^2 c^2}\: dp' \:,
\end{equation}
while for constant energy injection (equation~(\ref{eq:Rsw})), it is
written as
\begin{equation}
 K_Q = 16\pi^2 c\: \xi^{5-\omega}\left(\frac{3}{5-\omega}\right)^3
\frac{L_w}{E_{\rm cr,tot}}(t_f-t_0)
\int_{p_{\rm min}}^{p_{\rm max}}p'^{\: 2-q}
\sqrt{p'^{\: 2}+m_p^2 c^2}\: dp' \:, 
\end{equation}
where $c$ is the light velocity, and $m_p$ is the proton mass. CRs are
accelerated and injected into the Galactic halo space between $t=t_0$
and $t_f$, and $E_{\rm cr,tot}$ is the total energy of the CRs
accelerated during that period.  The maximum momentum $p_{\rm max}(t)$
is determined by the condition of $t_{\rm acc}=t_{\rm age}$, where
$t_{\rm acc}$ is the acceleration time-scale and $t_{\rm age}$ is the
age of the bubble. In our case, $t_{\rm age}=t$ and
\begin{equation}
\label{eq:pmax}
 p_{\rm max} \approx\frac{3}{20}\frac{e B_0}{\eta_g c^2}V_{\rm sh}^2 t\:,
\end{equation}
where $e$ is the proton charge, $B_0$ is the background magnetic field,
and $\eta_g$ is the gyro-factor \citep{aha99a,ohi10a}. We assume that
the shock is strong. Unless otherwise mentioned, we assume $\eta_g=1$
(Bohm diffusion). Instead of equation~(\ref{eq:pmax}), $p_{\rm max}$ is
often determined by an escape condition for SNRs
\citep{ptu05a,rev09}. This is because the characteristic spatial length
of particles penetrating into the shock upstream region can be
comparable to the size of SNRs \citep{ohi10a}. However, this cannot be
applied to the Fermi bubbles, because the size of the bubbles is much
larger than the characteristic length. We fix the minimum momentum at
$p_{\rm min}=m_p c$.

The CRs escaped from the shock neighborhood may amplify magnetic
fluctuations (Alfv\'en waves) in the Galactic halo through a streaming
instability \citep{wen74a,ski75c}. Since CRs are scattered by the
fluctuations, the diffusion coefficient $\kappa$ in
equation~(\ref{eq:diff}) is reduced.  At the rest frame and outside the
bubble ($r\geq R_{\rm sh}$), the wave growth is given by
\begin{equation}
\label{eq:psi}
 \frac{\partial\psi}{\partial t}
=\frac{4\pi}{3}\frac{v_A p^4 v}{U_M}|\nabla f|\:,
\end{equation}
where $\psi(r,p,t)$ is the energy density of Alfv\'en waves per unit
logarithmic bandwidth (which are resonant with particles of momentum
$p$) relative to the ambient magnetic energy density $U_M$
\citep{bel78a}, and $v$ is the particle velocity. The diffusion
coefficient is simply given as
\begin{equation}
\label{eq:kappa}
 \kappa(r,p,t) = \frac{4}{3\pi}\frac{p v c}{e B_0 \psi}\:.
\end{equation}
Within the bubble, the evolution of the waves could be complicated,
because it could be controlled by something like turbulence. Thus for
$r<R_{\rm sh}$, we simply assume that
\begin{equation}
\label{eq:kappa_b}
 \kappa(r,p,t)=\kappa_B\rho(R_{\rm sh})/\rho(r)\:,
\end{equation}
where $\kappa_B=\eta_g p c^2/(3 e B)$ is a Bohm-type diffusion
coefficient \citep{ber94a}, and $B\approx 4B_0$ for a strong shock. The
results do not much depend on the diffusion coefficient inside the
bubble if it is small enough. Although there is no strong
observational constraint on magnetic fields in the Galactic halo, we
assume that they are given by $B_0(r)=B_1(r/r_1)^{-\omega/2}$. For the
value of $B_1=10\:\rm \mu G$, the Alfv\'en velocity has a constant value
of $v_A=B_0/\sqrt{4\pi\rho}=100\rm\: km\: s^{-1}$.  The field
strength we assumed is comparable to or smaller than the value adopted
by \citet{su10a} or $B=30\: e^{-r/{\rm 2\: kpc}}\rm\: \mu G$ for
$r\lesssim 10$~kpc.  Gamma-ray and neutrino fluxes created through
hadronic interactions between CR protons and gas protons are calculated
using the code provided by \citet{kar08b}.

We consider four models. Model~FD is our fiducial model; an energy of
$E_{\rm tot}=2.5\times 10^{57}$~erg is instantaneously released at $t=0$
at $r=0$ (see equation~(\ref{eq:Rs})). For the initial distribution of
the halo gas, we assume that $\omega=1.5$, $\rho_1=7.8\times
10^{-24}\rm\: g\: cm^{-3}$ and $r_1=0.1$~kpc to be consistent with
gamma-ray observations (Section~\ref{sec:result}). The index
$\omega=1.5$ is an approximation of the profile obtained by
\citet{guo12a}. We solve equations~(\ref{eq:diff}) and~(\ref{eq:psi})
for $t\geq t_0=1\times 10^6$~yr. It is to be noted that if the SMBH at
the GC injects energy at a rate of 16\% of the Eddington luminosity
($\sim 5\times 10^{44}\rm\: erg\: s^{-1}$) for $1\times 10^6$~yr, the
total energy is comparable to $E_{\rm tot}$. The energy that goes into
CRs is $E_{\rm cr,tot}=0.2\: E_{\rm tot}$ in total. The spectral index
of the accelerated CRs is assumed to be $q=4.1$. For the parameters we
adopted, the maximum momentum at $t=t_0$ is $p_{\rm max}c=9\times
10^{14}$~eV. At $t=t_0$, we assume that the diffusion coefficient has
typical Galactic values:
\begin{equation}
\label{eq:kappa_i}
 \kappa_i=10^{28}\left(\frac{E}{10\rm\: GeV}\right)^{0.5}
\left(\frac{B_0}{3\: \mu\rm G}\right)^{-0.5}\rm\: cm^2\: s^{-1}\:,
\end{equation}
where $E$ is the energy of a CR proton \citep{gab09a}. From
equation~(\ref{eq:kappa}), we obtain the initial wave energy density
$\psi_i(r,p)=\psi(r,p,t_0)\propto \kappa_i^{-1}$.  If the temperature of
the halo gas is $T=2.4\times 10^6$~K \citep{guo12a}, the Mach number of
the shock at $t=3\times 10^6$~yr is ${\cal M}\sim 4$. Since it is
generally believed that CR acceleration is ineffective at smaller Mach
numbers \citep[e.g.][]{gie00a}, we assume that CR acceleration finishes
at $t_f=3\times 10^6$~yr. The current age of the bubble is assumed to be
$t_{\rm obs}=10^7$~yr and the bubble center (GC) is located at a
distance of 8.5~kpc. The current bubble size is $R_{\rm sh}(t_{\rm
obs})=9.7$~kpc.

Other models are studied for comparison. Their parameters are the same
as those for Model~FD except for the followings. In Model~NG, the wave
growth is ignored, and we assume that $\kappa=\kappa_i$ for $r>R_{\rm
sh}$. In Model~LA, acceleration of CRs starts later, and we adopt
$t_0=4\times 10^6$~yr and $t_f=t_{\rm obs}=10^7$~yr. In Model~CI, energy
is continuously injected from the GC at a rate of $L_w=E_{\rm
tot}/t_{\rm obs}$ for $t<t_{\rm obs}$ (see equation~(\ref{eq:Rsw})), and
we set $t_{\rm obs}=2\times 10^7$~yr so that the position of the peak of
the surface brightness profiles is almost the same as that of Model~FD.

\section{Results}
\label{sec:result}

\begin{figure}
\epsscale{.80} \plotone{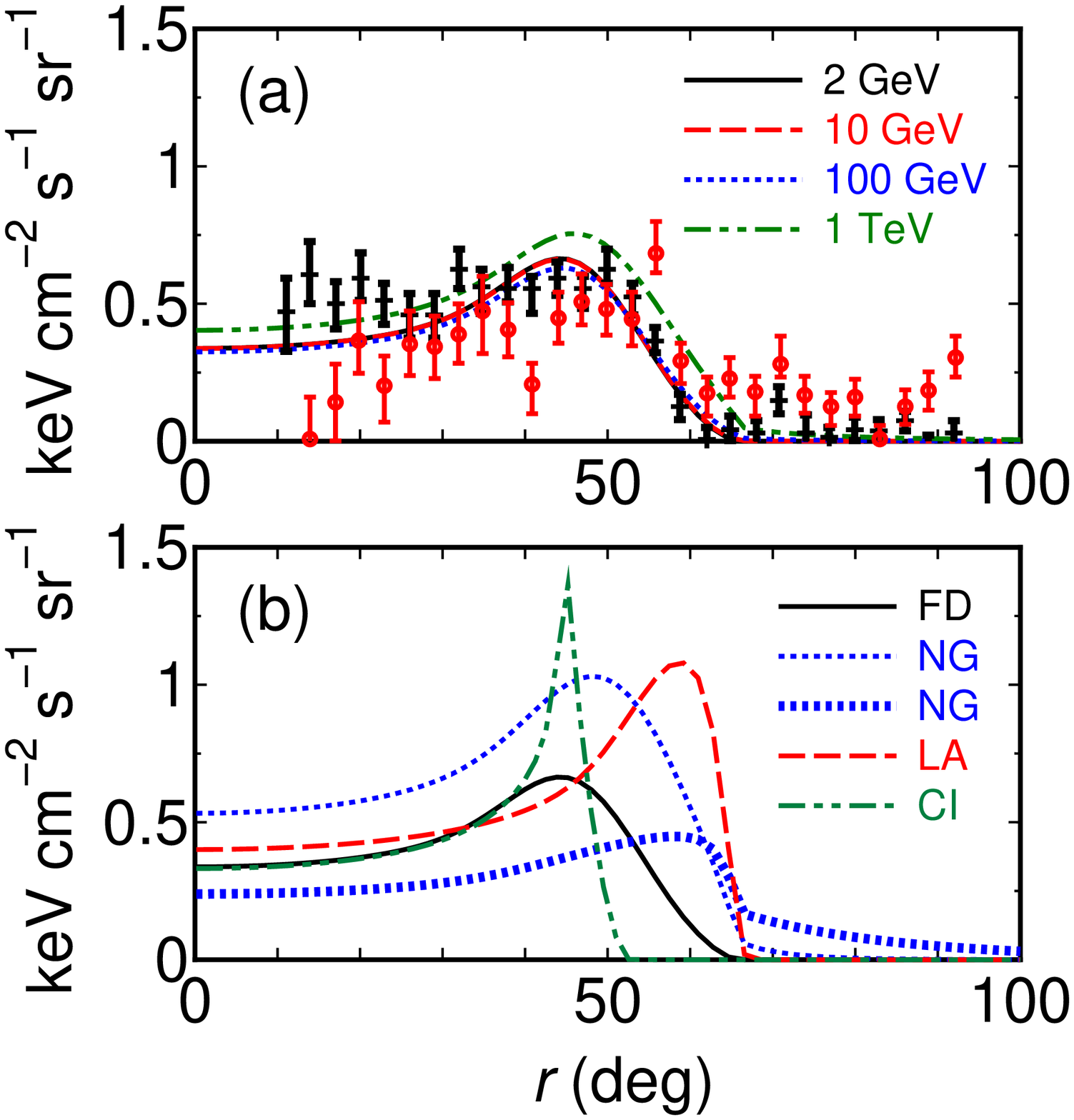} \caption{(a) Profiles of gamma-ray
surface brightness at 2~GeV (solid), 10~GeV (dotted), 100~GeV (dashed),
and 1~TeV (two-dotted dashed) for Model~FD. Crosses and circles are the
observations for the southern bubble by \citet{su10a} in the 1--5~GeV
band (cross) and 5--20~GeV band (circle), respectively. (b) Profiles of
gamma-ray surface brightness at 2~GeV for Model~FD (solid), at 2~GeV for
Model~NG (thin dotted), at 1~TeV for Model~NG (thick dotted), at 2~GeV
for Model~LA (dashed), and at 2~GeV for Model~CI (two-dotted
dashed). For comparison, the surface brightness is multiplied by 0.5 for
Model~NG (2~GeV) and Model~LA.}\label{fig:pro}
\end{figure}

\begin{figure}
\epsscale{.80} \plotone{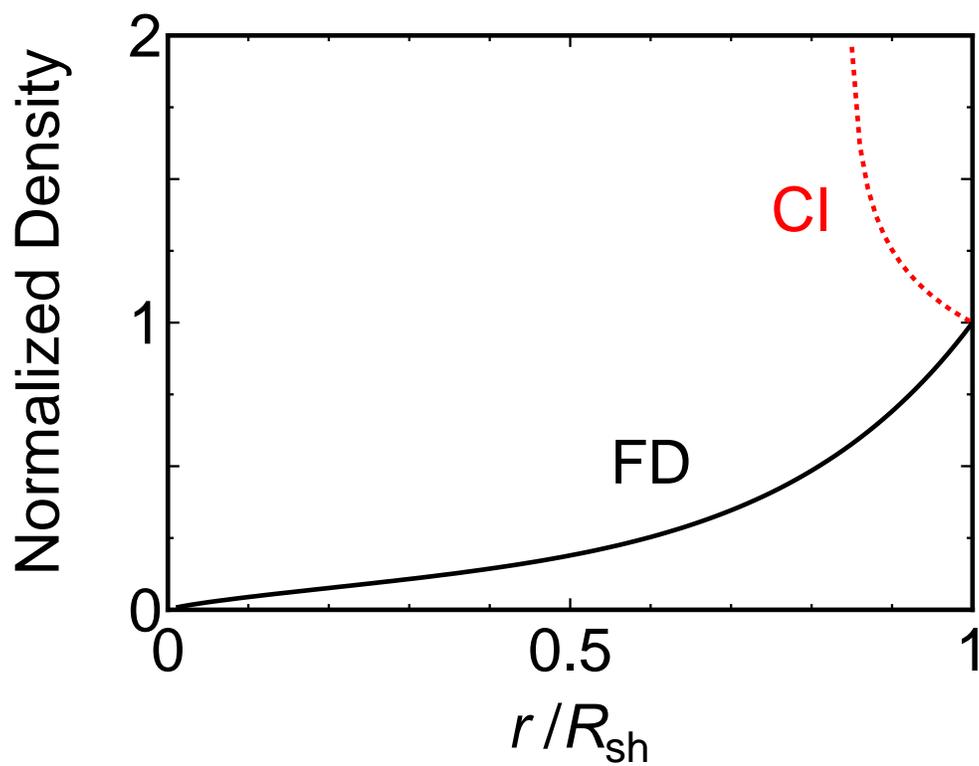} \caption{Gas density profiles within the
bubble for Model~FD (solid) and Model~CI (dotted). The density is
normalized by the value at $r=R_{\rm sh}$.}\label{fig:sedov}
\end{figure}

Figure~\ref{fig:pro} shows the surface brightness profiles of the
bubble. They are calculated simply by projecting gamma-ray emissions on
a plane at a distance of 8.5~kpc and we do not consider the detailed
geometrical effects that come up when the distance to the Fermi bubbles
is finite. Figure~\ref{fig:pro}a shows the results for Model~FD, which
are compared with the southern bubble data in Figure~9 in \citet{su10a};
one degree corresponds to $\pi/180\times 8.5$~kpc. Since our model is
rather simple and we do not include background, we shift the
observational data along the horizontal axis ($+5^\circ$) and the
vertical axis ($-0.9\rm\: keV\: cm^{-2}\: s^{-1}\: sr^{-1}$ in the 1--5
GeV band and $-0.4\rm\: keV\: cm^{-2}\: s^{-1}\: sr^{-1}$ in the 5--20
band). At 2 and 10~GeV, the predicted profiles are fairly flat and have
sharp edges at $r\sim 50^\circ$ as observations suggest
(Figure~\ref{fig:pro}a). Significant gamma-ray emissions fill the
bubble, because not all the gas is concentrated at the shock. (Model~FD
in Figure~\ref{fig:sedov}).

\begin{figure}
\epsscale{.80} \plotone{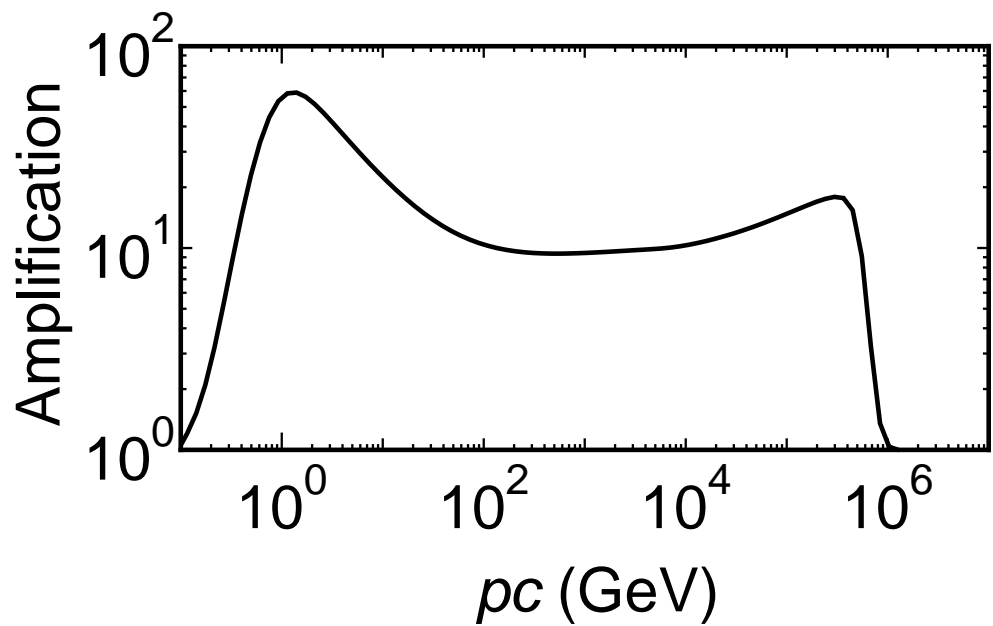} \caption{Amplification of magnetic
fluctuations $\psi/\psi_i$ at $r=R_{\rm sh,+}$ at $t=t_{\rm obs}$ for
Model~FD.}  \label{fig:amp}
\end{figure}

The sharp edges seen in Figure~\ref{fig:pro}a are created by the dense
gas just behind the shock. Moreover, the wave amplification outside the
shock also contributes to the formation of the sharp
edges. Figure~\ref{fig:amp} shows the wave amplification $\psi/\psi_i$
at $r=R_{\rm sh,+}$ and $t=t_{\rm obs}$ for Model~FD. The amplification
leads to the decrease of the diffusion coefficient
(equation~(\ref{eq:kappa})) and slows the CR diffusion out of the
bubble. While the wave energy $\psi$ grows at a given radius outside the
shock (equation~(\ref{eq:psi})), the wave energy at the expanding shock
($r=R_{\rm sh,+}(t)$) gradually decreases.  While $\psi/\psi_i$ is not
much dependent on CR momentum at $10^{10}\lesssim pc \lesssim
10^{14}$~eV (Figure~\ref{fig:amp}), $\psi_i\propto \kappa_i^{-1}$ is a
decreasing function of CR momentum (equations~(\ref{eq:kappa}) and
(\ref{eq:kappa_i})). This means that $\kappa\propto\psi^{-1}$ is an
increasing function of CR momentum and CRs with larger energies diffuse
faster. Assuming that CR acceleration stopped at $t=t_f < t_{\rm obs}$,
CRs are left far behind the shock at $t=t_{\rm obs}$ if their diffusion
is not much effective. This happens for GeV CRs in Model~FD; most of
them remain far behind the shock. However, this is not the case for CRs
with much larger energies. At $t=t_{\rm obs}$ and $r=R_{\rm sh,+}$, the
diffusion coefficient for CRs with $pc=10$~TeV is $\kappa=9.7\times
10^{28}\rm\: cm^2\: s^{-1}$. Thus, the diffusion scale-length is $l_{\rm
diff}\sim\sqrt{4\kappa (t_{\rm obs}-t_f)}\sim 3.0$~kpc. On the other
hand, the shock velocity at $t=t_{\rm obs}$ is $V_{\rm sh}=540\rm\: km\:
s^{-1}$, and thus the advection scale-length is $l_{\rm adv}=V_{\rm
sh}(t_{\rm obs}-t_f)\sim 3.9$~kpc. Since $l_{\rm diff}\lesssim l_{\rm
adv}$, most CRs do not diffuse beyond the shock radius, although the
diffusion cannot be ignored. This explains the profile of 1~TeV
gamma-rays, which is created by CRs with $\sim 10$~TeV
(Figure~\ref{fig:pro}a). The moderate diffusion enables some of the TeV
CRs to reach the very high density region just behind the shock. This
makes the surface brightness at 1~TeV a little brighter than those at
smaller energies (Figure~\ref{fig:pro}a). We note that the slight
increase of $\psi/\psi_i$ at $pc\sim 10^{14}$~eV in Figure~\ref{fig:amp}
is caused by the higher-energy CRs that have arrived at $r\sim R_{\rm
sh}$. In Model~NG, in which the wave growth is ignored, the diffusion
for GeV CRs is the moderate one and the surface brightness in the GeV
band is larger than that for Model~FD (Figure~\ref{fig:pro}b). However,
the diffusion of CRs with $\gtrsim$~TeV is much faster and the CRs
diffuse beyond the shock radius (Figure~\ref{fig:pro}b). While the
surface brightness profile for a given energy is relatively flat for
Model~NG, the spatial extension of CRs varies with their energies
because of the fairly large and energy-dependent diffusion
(equation~(\ref{eq:kappa_i})).

In Model~LA, the limb-brightening becomes more prominent because CRs do
not have much time to diffuse out (Figure~\ref{fig:pro}b). Thus, $t_0$
must be much smaller than $t_{\rm obs}$, or CR acceleration must have
started at the shock when the bubble is small. The observed flatness of
the surface brightness profiles may also imply that the time-scale of
the energy injection at the GC is much shorter than the age of the
bubble ($t_{\rm inj}\ll t_{\rm obs}$). In Models~FD, NG and LA, we
implicitly assumed that $t_{\rm inj}\lesssim t_0$. On the other hand, in
Model~CI the energy has been continuously injected at the GC ($t_{\rm
inj}=t_{\rm obs}$). In this case, the halo gas inside the bubble
($r<R_{\rm sh}$) is compressed into a thin dense shell between the shock
and the contact discontinuity at $r=0.86\: R_{\rm sh}$
(Figure~\ref{fig:sedov}). If the region behind the contact discontinuity
($r<0.86\: R_{\rm sh}$) is almost empty with gas, the gamma-ray image
should have a shell-like structure (Model~CI in Figure~\ref{fig:pro}b),
because gamma-rays are created through the interaction between CRs and
the gas of the thin shell. In Model~CI, we assume that the gas density
inside the contact discontinuity is $0.1\:\rho(R_{\rm sh,+})$ for a
calculational purpose. 

\begin{figure}
\epsscale{.80} \plotone{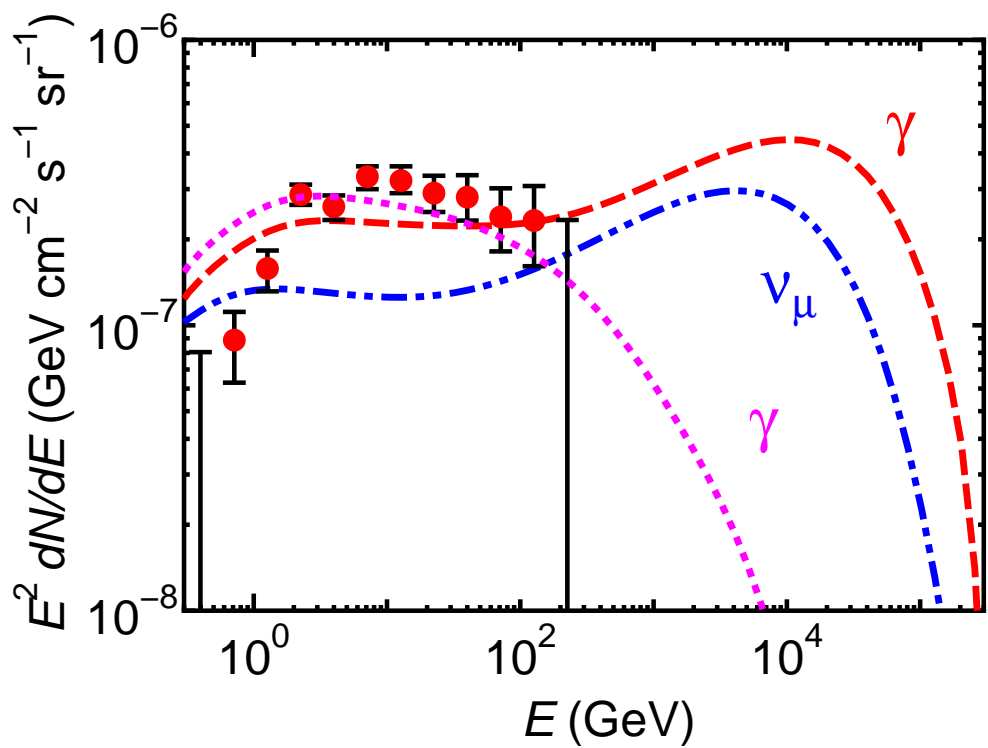} \caption{Gamma-ray (dashed) and neutrino
($\nu_\mu+\bar{\nu}_\mu$; two-dotted dashed) fluxes for Model~FD. Dotted
line is the gamma-ray spectrum when $p_{\rm max}$ is small (see
text). Filled circles are the observations \citep{su12b}.}
\label{fig:sp}
\end{figure}

Figure~\ref{fig:sp} shows the gamma-ray spectrum of the bubble for
Model~FD. The observed hard gamma-ray spectrum is reproduced, which
reflects that the spectral index of the CRs around the bubble is not
much different from the original one ($q=4.1$). This is because of the
decrease of the diffusion coefficient or the confinement of CRs around
the bubble. For Model~NG, the confinement depends on CR energies, and
the original CR spectrum is not conserved \citep{ohi11b}. In
Figure~\ref{fig:sp}, the gamma-ray luminosity in the TeV band is
slightly larger than that in the GeV band as was shown in
Figure~\ref{fig:pro}a. We note that the TeV luminosity depends on
$p_{\rm max}$. For example, larger $\eta_g$ makes $p_{\rm max}$ smaller
(equation~(\ref{eq:pmax})). The dotted line in Figure~\ref{fig:sp} shows
the spectrum when $\eta_g=100$; other parameters are the same as those
for Model~FD. As can be seen, the TeV luminosity is much reduced. In
Figure~\ref{fig:sp}, we also present the neutrino spectrum for
Model~FD. The flux is similar to the one predicted by \citet{lun12a},
and thus their discussion can be applied. Our results indicate that the
neutrino flux is comparable to the background and it could be marginally
detected \citep{lun12a}.

Model~FD indicates that the position of the shock is a few kpc outside
the edge of the gamma-ray bubble. X-ray emission from the high-density
gas just behind the shock could have been detected there
\citep{sof00a,bla03a}. In Model~FD, in which CRs are accelerated to PeV,
we also predict that the size of the bubble is larger in the TeV band
(Figure~\ref{fig:pro}a), because of the faster diffusion of higher
energy CRs. The difference of the size of the bubble between the GeV and
TeV bands could be used to discriminate our hadronic model from other
leptonic models. In the leptonic models, gamma-rays originate from
electrons and the cooling time of the electrons with energies of
$\gtrsim 30$~GeV is smaller than the age of the bubbles \citep[Figure~28
in][]{su10a}. This means that the electrons must be being
accelerated. Thus, the gamma-rays should be observed around the
acceleration sites and the gamma-ray distribution should not depend on
the energy band. In particular, since the cooling time of TeV electrons
is very short ($\lesssim 10^6$~yr), the diffusion of the electrons can
be ignored and the distribution of the gamma-rays from them should
reflect the positions of the acceleration sites.

\section{Summary and Discussion}

We solve a diffusion-advection equation to investigate the evolution of
the distribution of CRs accelerated at the shock front of the Fermi
bubbles. We found that the observed flat surface brightness profile with
a sharp edge can be reproduced because of the gas inside the bubbles and
the reduced diffusion coefficient owing to CR streaming
instabilities. The latter also contributes to the hard spectrum of the
bubbles. The CR acceleration must have started at the early stage of the
bubble evolution and the time-scale of energy injection at the GC must
be much smaller than the current age of the bubbles.

the hadronic model by \citet{cro11a}, the bubbles are long-lived
($\gtrsim 8$~Gyr) or steady. This is not likely in our model because the
forward shocks rapidly cross the halo ($\sim 10^7$~yr), unless the
background gas is rapidly falling toward the galactic plane. If the SMBH
blows winds, reverse shocks may develop in the winds \citep*{zub11a},
and CRs may be accelerated there. Our model does not treat this type of
acceleration. If the reverse shocks disappear in a short time, the
situation may not be much different from the one we considered.
Moreover, since the reverse shocks are located in the innermost region
of the bubbles, it may take a longer time for the CRs accelerated there
to diffuse out to the dense region at the bubble edge than those
accelerated at the forward shocks. Thus, the contribution of the former
to the gamma-ray emission may be less than that of the
latter. \citet{zub12a} indicated that the activity of the SMBH lasted
$\sim 10^6$~yr and the current age of the bubbles is $\sim 10^7$~yr,
which are consistent with our model. They also indicated that buoyancy
may deform the bubbles. Although it may somewhat change the height of
the bubbles, our results may not be much affected as long as CRs
generally move with the background gas.

\acknowledgments

This work was supported by KAKENHI (YF: 23540308, YO:
No.24.8344). R.~Y. was supported by the fund from Research Institute,
Aoyama Gakuin University.

\clearpage

\begin{thebibliography}{}

\bibitem[Aharonian 
\& Atoyan(1999)]{aha99a} Aharonian, F.~A., \& Atoyan, A.~M.\ 1999, \aap,
		351, 330

\bibitem[Bell(1978)]{bel78a} Bell, A.~R.\ 1978, \mnras, 182, 
147 

\bibitem[Bell(2004)]{bel04a} Bell, A.~R.\ 2004, \mnras, 353,
550 

\bibitem[Berezhko et al.(1994)Berezhko, Yelshin \& Ksenofontov]{ber94a}
		Berezhko, E.~G.,
Yelshin, V.~K., \& Ksenofontov, L.~T.\ 1994, Astroparticle Physics, 2,
		215

\bibitem[Bland-Hawthorn 
\& Cohen(2003)]{bla03a} Bland-Hawthorn, J., \& Cohen, M.\ 2003, \apj,
		582, 246

\bibitem[Cheng et al.(2011)]{che11a} Cheng, K.-S., Chernyshov, 
D.~O., Dogiel, V.~A., Ko, C.-M., \& Ip, W.-H.\ 2011, \apjl, 731, L17 

\bibitem[Crocker 
\& Aharonian(2011)]{cro11a} Crocker, R.~M., \& Aharonian, F.\ 2011,
		Physical Review Letters, 106, 101102

\bibitem[Crocker(2012)]{cro12a} Crocker, R.~M.\ 2012, \mnras, 
423, 3512 

\bibitem[Dobler et al.(2010)]{dob10} Dobler, G., Finkbeiner, 
D.~P., Cholis, I., Slatyer, T., \& Weiner, N.\ 2010, \apj, 717, 825 

\bibitem[Drury(1983)]{dru83a} Drury, L.~O.\ 1983, Reports on 
Progress in Physics, 46, 973 

\bibitem[Fujita et al.(2010)Fujita, Ohira, 
\& Takahara]{fuj10a} Fujita, Y., Ohira, Y., 
\& Takahara, F.\ 2010, \apjl, 712, L153 

\bibitem[Fujita et al.(2009)]{fuj09c} Fujita, Y., Ohira, Y., 
Tanaka, S.~J., \& Takahara, F.\ 2009, \apjl, 707, L179 

\bibitem[Fujita et al.(2011)]{fuj11a} Fujita, Y., Takahara, 
F., Ohira, Y., \& Iwasaki, K.\ 2011, \mnras, 415, 3434 

\bibitem[Gabici et al.(2009)Gabici, Aharonian, 
\& Casanova]{gab09a} Gabici, S., Aharonian, 
F.~A., \& Casanova, S.\ 2009, \mnras, 396, 1629

\bibitem[Gieseler et 
al.(2000)]{gie00a} Gieseler, U.~D.~J., Jones, T.~W., \& Kang, H.\ 2000,
		\aap, 364, 911

\bibitem[Guo 
\& Mathews(2012)]{guo12a} Guo, F., \& Mathews, W.~G.\ 2012, \apj, 756,
		181

\bibitem[Guo et al.(2012)]{guo12b} Guo, F., Mathews, W.~G., 
Dobler, G., \& Oh, S.~P.\ 2012, \apj, 756, 182 

\bibitem[Karlsson 
\& Kamae(2008)]{kar08b} Karlsson, N., \& Kamae, T.\ 2008, \apj, 674, 278


\bibitem[Li 
\& Chen(2010)]{li10a} Li, H., \& Chen, Y.\ 2010, \mnras,
	      409, L35 

\bibitem[Lucek 
\& Bell(2000)]{luc00a} Lucek, S.~G., \& Bell, A.~R.\ 2000, \mnras, 314,
		65

\bibitem[Lunardini 
\& Razzaque(2012)]{lun12a} Lunardini, C., \& Razzaque, S.\ 2012,
		Physical Review Letters, 108, 221102


\bibitem[Malkov et al.(2013)]{mar13a} Malkov, M.~A., Diamond,
P.~H., Sagdeev, R.~Z., Aharonian, F.~A., 
\& Moskalenko, I.~V.\ 2013, \apj, 768, 73 

\bibitem[Mertsch 
\& Sarkar(2011)]{mer11a} Mertsch, P., \& Sarkar, S.\ 2011, Physical
		Review Letters, 107, 091101


\bibitem[Mihalas \& Mihalas(1984)]{mih84} Mihalas, D., \& Mihalas,
		B. W.\ 1984, Foundations of Radiation Hydrodynamics (New
		York: Oxford University Press), \S~60


\bibitem[Nava 
\& Gabici(2013)]{nav13a} Nava, L., \& Gabici, S.\ 2013, \mnras, 429,
		1643

\bibitem[Ohira et 
al.(2010)Ohira, Murase, \& Yamazaki]{ohi10a} Ohira, Y., Murase, K., \&
		Yamazaki, R.\
	  2010, \aap, 513, A17 

\bibitem[Ohira et al.(2011)Ohira, Murase, \& Yamazaki]{ohi11b} Ohira,
		Y., Murase, K.,
\& Yamazaki, R.\ 2011, \mnras, 410, 1577 

\bibitem[Ostriker \& McKee(1988)]{ost88} Ostriker, J.~P., \& 
McKee, C.~F.\ 1988, Reviews of Modern Physics, 60, 1 

\bibitem[Ptuskin 
\& Zirakashvili(2005)]{ptu05a} Ptuskin, V.~S., \& Zirakashvili, V.~N.\
		2005, \aap, 429, 755


\bibitem[Reville et al.(2009)Reville, Kirk, \& Duffy]{rev09} Reville,
		B., Kirk,
J.~G., \& Duffy, P.\ 2009, \apj, 694, 951 

\bibitem[Skilling(1975)]{ski75c} Skilling, J.\ 1975, \mnras, 
173, 255 

\bibitem[Sofue(2000)]{sof00a} Sofue, Y.\ 2000, \apj, 540, 224 

\bibitem[Su 
\& Finkbeiner(2012)]{su12b} Su, M., \& Finkbeiner, D.~P.\ 2012, \apj,
		753, 61

\bibitem[Su et al.(2010)Su, Slatyer,
\& Finkbeiner]{su10a} Su, M., Slatyer, T.~R., 
\& Finkbeiner, D.~P.\ 2010, \apj, 724, 1044 

\bibitem[Torres et al.(2008)]{tor08b} Torres, D.~F., Rodriguez 
Marrero, A.~Y., \& de Cea Del Pozo, E.\ 2008, \mnras, 387, L59 

\bibitem[Wentzel(1974)]{wen74a} Wentzel, D.~G.\ 1974,
		\araa, 12, 71


\bibitem[Yan et al.(2012)Yan, Lazarian,
\& Schlickeiser]{yan12b} Yan, H., Lazarian, A., 
\& Schlickeiser, R.\ 2012, \apj, 745, 140 


\bibitem[Yang et al.(2012)]{yan12a} Yang, H.-Y.~K., 
Ruszkowski, M., Ricker, P.~M., Zweibel, E., 
\& Lee, D.\ 2012, \apj, 761, 185 

\bibitem[Zubovas et al.(2011)Zubovas, King, \& Nayakshin]{zub11a}
Zubovas, K., King, A.~R., \& Nayakshin, S.\ 2011, \mnras, 415, L21

\bibitem[Zubovas 
\& Nayakshin(2012)]{zub12a} Zubovas, K., \& Nayakshin, S.\ 2012, \mnras,
		424, 666


\end{thebibliography}
\end{document}